\documentstyle[twocolumn,aps,epsf]{revtex}

\def\i{{\rm i}}
\def\d{{\rm d}}

\def\vector#1{{\bf #1}}

\def\vk{{\vector k}}
\def\vq{{\vector q}}

\def\vr{{\vector r}}

\def\dps{\displaystyle}

\def\Tc{{T_{\rm c}}}

\def\qbar{{\bar q}}
\def\hightc{{high-$T_{\rm c}$ }} 

\def\nh{{n_{\rm h}}}

\def\kB{{k_{\rm B}}}

\def\hsp#1{\hspace{#1ex}}

\def\Tc{{T_{\rm c}}}

\def\lsim{\stackrel{{\textstyle<}}{\raisebox{-.75ex}{$\sim$}}}

\def\parallelsl{{\!/\!\!/}}

\def\kF{k_{{\rm F}}}

\begin{document}
\draft

\twocolumn[\hsize\textwidth\columnwidth\hsize\csname 
@twocolumnfalse\endcsname

\title{
Temperature dependence of the upper critical field of an anisotropic \\
singlet superconductivity in a square lattice tight-binding model \\
in parallel magnetic fields 
}

\author{Hiroshi Shimahara and Kentaro Moriwake}

\def\runtitle{Temperature dependence of the upper critical field 
of superconductors on the square lattices 
in parallel magnetic field 
}

\def\runauthor{Hiroshi {\sc Shimahara} and Kentaro {\sc Moriwake}}

\address{
Department of Quantum Matter Science, ADSM, Hiroshima University, 
Higashi-Hiroshima 739-8530, Japan
}

\date{January 26, 2002} 

\maketitle

\begin{abstract}
Upper critical field parallel to the conducting layer 
is studied in anisotropic type-II superconductors 
on square lattices. 
We assume enough separation of the adjacent layers, 
for which the orbital pair-breaking effect is suppressed 
for exactly aligned parallel magnetic field. 
In particular, we examine the temperature dependence of the critical 
field $H_{c}(T)$ of the superconductivity 
including the Fulde-Ferrell-Larkin-Ovchinnikov (FFLO or LOFF) state, 
in which the Cooper pairs have non-zero center-of-mass momentum $\vq$. 
In the system with the cylindrically symmetric Fermi-surface, 
it is known that $H_{\rm c}(T)$ of the $d$-wave FFLO state 
exhibits a kink at a low temperature 
due to a change of the direction of $\vq$ 
in contrast to observations in organic superconductors. 
It is shown that the kink disappears when the Fermi-surface is anisotropic 
to some extent, 
since the direction of $\vq$ is locked in an optimum direction 
independent of the temperature. 
\end{abstract}

\pacs{
}


]

\narrowtext

Recently, upturn of the curve of the upper critical field as a function of 
the temperature has been observed in organic superconductors 
when the magnetic field is applied in the directions parallel to 
the conducting layers~\cite{Ish00,Sin00,Ohm99,Tan01,Ohm01,Lee97}. 
This behavior has often been 
discussed~\cite{Shi94,Bur94,Shi99,Shi97a,Shi97b,Man00,Shi00b}
in connection with a possibility of 
the Fulde-Ferrell-Larkin-Ovchinnikov (FFLO) state~\cite{Ful64,Lar64}. 
It is known by many theoretical calculations~\cite{Bul73,Aoi74,Shi94,Bur94,Shi99} 
that the parallel critical field exhibits upturn at low temperatures 
for the FFLO state in some models. 
However, in $d$-wave pairing, 
the curve of the critical field has a kink above the temperatures 
at which the upturn occurs when the orbital pair-breaking effect is 
negligible~\cite{Mak96,Shi97b,Yan98,Shi00b}, 
althought the kink has not been observed yet. 
In this paper, we show that the kink is suppressed by anisotropy of 
the Fermi-surfaces of some extent.

Firstly, 
we briefly review theories and experiments which are concerned with 
the present theory. 
The superconductivity in type II superconductor is suppressed by applied 
magnetic fields due to both the Lorentz force and Zeeman energy 
for the singlet pairing. 
However, the Lorentz force is not effective, 
when the magnetic field is applied in directions parallel to 
the conducting layers, 
since the spatial motion of electron is restricted 
in the conducting layers. 
In this situation, the pair-breaking is mainly due to the Zeeman energy, 
and the upper critical field is known as the Chandrasekhar and Clogston 
limit (Pauli paramagnetic limit)~\cite{Cha62,Clo62}.

For example, for $s$-wave pairing, 
the critical field at zero temperature 
$H_{\rm P}^{(0)}$ 
is estimated from the zero field transition temperature 
$T_{\rm c}^{(0)}$ by a formula 
$\mu_0 H_{\rm P}^{(0)} \approx 1.25 \times \kB T_{\rm c}^{(0)}$ 
in the weak coupling limit, 
where $\mu_0$ is the electron magnetic moment. 
The critical field is maximum at $T = 0$ 
with $[\d H_{\rm P}(T)/\d T]_{T = 0} = 0$, 
and it is upwards convex at low temperature 
($\d^2 H_{\rm P}(T)/\d T^2 < 0$).

Fulde and Ferrell~\cite{Ful64}, 
and independently Larkin and Ovchinnikov~\cite{Lar64} 
predicted that the critical field can exceed 
the original Pauli paramagnetic limit 
by a pairing state with a non-zero 
center-of-mass momentum $\vq$. 
This state is called the FFLO or LOFF state, 
and as a direct consequence of the non-zero center-of-mass momentum, 
the superconducting order parameter oscillates in space. 
Larkin and Ovchinnikov found that the order parameter has 
the spatial structure like $\cos(\vq \cdot \vr)$ 
near the critical field in three dimensions 
rather than the structure like $\exp(\i \vq \cdot \vr)$ 
by calculations of the free energies of the states, 
although both structures have the same upper critical field 
within the second order transition.

The critical field of the FFLO state in three dimensions has 
a similar temperature dependence to 
the Pauli limit as mentioned above, 
wherease the value is slightly larger than the Pauli limit. 
Furthermore, the orbital pair-breaking effect strongly 
suppresses the FFLO state. 
The condition for the coexistence of the FFLO state and the vortex 
state was examined by Gruenberg and Gunther~\cite{Gru66}. 
When the FFLO state coexists with the vortex state, 
the spatial oscillation of the order parameter due to the FFLO state 
is along the vortex lines.

On the other hand, in the two dimensions, 
the critical field as a function of the temperature is downwards convex 
($\d^2 H_{\rm c2}(T)/\d T^2 > 0$) 
at low temperatures~\cite{Bul73,Bur94,Shi94}. 
This behavior, which is called upturn, 
is due to a Fermi-surface effect analogous to 
nesting effects of spin and charge density 
waves~\cite{Shi94,Shi99,Shi97a,Shi00a}. 
In the two dimensions, the Fermi-surfaces of 
up and down spin electrons touch on lines by a shift by $\vq$ 
at $T = 0$. 
It was found that by a calculation of the free energy 
that the order parameter has various two dimensional structures 
at low temperatures and high fields 
in the two dimensions~\cite{Shi98}. 
The origin of the two dimensional structure is the high critical field, 
for which a gain in the polarization energy is more important 
than a loss in the condensation energy.

Temperature dependences similar to the theoretical prediction 
including the upturn have been observed in the organic 
superconductors~\cite{Sin00,Ish00,Ohm99,Tan01,Ohm01,Lee97}. 
Hence, the FFLO state has often been discussed as a candidate 
to explain the critical field, 
although dimensional crossover effects in triplet and singlet 
superconductors are also possible mechanisms of 
the upturn~\cite{Leb99,Miy99}. 
However, the observation of the lower critical field 
in $\kappa$-${\rm (BEDT}$-${\rm TTF)}_2{\rm Cu(NCS)}_2$~\cite{Sin00}, 
a change in the temperature dependence of upper critical field 
due to the purity of the samples of 
$\lambda$-${\rm (BETS)_2GaCl_4}$~\cite{Tan01}, 
and that due to tilt angle of the magnetic field 
in $\kappa$-${\rm (BEDT}$-${\rm TTF)}_2{\rm Cu[N(CN)_2]Br}
$~\cite{Ohm01,Man00,Shi97b} 
seem to support the FFLO state in these organic superconductors. 
In the phase diagram of $\lambda$-${\rm (BETS)_2FeCl_4}$, 
the lower critical field of the field induced superconductivity 
shows upturnlike behavior ($\d^2 H_{\rm c2}(T)/\d T^2 < 0$) 
which is naturally explained by the FFLO state in combination 
with the Jaccarino and Peter mechanism~\cite{Uji01,Jac62,Shi02}.

However, when we consider the FFLO state in the organic superconductors 
and the \hightc cuprates, 
we should note that $d$-wave pairing is an important candidate 
of the pairing. 
The upturn of the critical field occurs also for the $d$-wave pairing 
as studied by Maki and Won~\cite{Mak96} 
and many other authors~\cite{Shi97b,Yan98,Shi00b}. 
However, in addition to the upturn at low temperatures 
($T \lsim 0.06 T_{\rm c}^{(0)}$), 
the upper critical field $H_{\rm c}(T)$ exhibits a kink 
at $T \approx 0.06 T_{\rm c}^{(0)}$ 
due to a change in the direction 
of $\vq$~\cite{Mak96,Shi97b,Yan98,Shi00b}. 
When we assume $d_{x^2-y^2}$-wave pairing, 
$\vq$ is oriented to one of the crystal axes 
($\parallelsl x$-axis or $\parallelsl y$-axis) 
for lower temperatures, 
whereas it is oriented to the direction of $(1,1,0)$ 
for higher temperatures 
($0.06 T_{\rm c}^{(0)} \lsim T \lsim 0.56 T_{\rm c}^{(0)}$). 
This behavior does not seem to be consistent with the experimental data 
of the upper critical fields~\cite{Sin00,Ish00,Ohm99,Tan01,Ohm01,Lee97}.

Here, we should add a discussion for the case in which 
the vortex state coexists with the FFLO state 
in the presence of the orbital pair-breaking effect. 
In this case, 
$\vq$ of the FFLO state is fixed in the direction of the vortex line, 
and hence the kink does not occur. 
However, 
the temperature dependence of the critical field 
would change completely 
depending on the direction of the magnetic field within the directions 
parallel to the layers.

In this paper, we examine the case in which the orbital pair-breaking 
effect is negligible and $\vq$ is freely oriented to the optimum direction 
at each temperature and field. 
The orbital pair-breaking effect is negligible, 
when separation of adjacent layers is sufficiently large~\cite{Leb99}, 
the layers are sufficiently thin~\cite{Man02}, 
and the magnetic field is exactly aligned in a direction parallel to 
the layers~\cite{Shi97b,Shi99}. 
It is shown that anisotropy of the Fermi-surface 
locks the optimum direction of $\vq$, 
and thus the kink of $H_{\rm c}(T)$ disappears. 
We examine a tight-binding model on the square lattice~\cite{Shi99,Miy99} 
as an example of an anisotropic system. 
By change of the hole concentration, 
the shape of the Fermi-surface can be controlled 
between square and circular shapes.

At zero temperature, the hole concentration dependence of the 
critical field has been examined 
in the square lattice tight-binding model~\cite{Shi99}. 
At hole concentration per a site $\nh \approx 0.630$, 
(i.e., electron density per a site $n \approx 0.370$), 
the nesting is most effective, 
and the critical field is extraordinarily enhanced 
near this hole concentration. 
The Pauli paramagnetic limit $H_{\rm P}^{(0)}$ 
was also calculated in this paper, 
and it was found that the FFLO critical field exceeds $H_{\rm P}^{(0)}$ 
at all hole concentrations 
except in a small region near the half-filling.

In the present theory, we concentrate ourselves on the temperature 
dependence of the upper critical field. 
We consider sufficiently weak coupling of the pairing interaction 
and assume implicite small inter-layer electron hopping, 
which makes the mean field treatment appropriate.

The gap equation that we examine is 
\def\eqgapeq{(1)}
$$
     \Delta_{\vq} = \frac{V}{N} \sum_{\vk} 
     [\gamma_{\alpha}(\vk)]^2 
     \frac{1 - f(E_{\vk \uparrow}) - f(E_{\vk \downarrow})} 
          {2 E_{\vk}} 
     \Delta_{\vq} , 
     \eqno\eqgapeq
     $$
where $f(\epsilon)$ is the Fermi distribution function, 
$\gamma_{\alpha}(\vk)$ is a symmetry factor of the gap function, 
$E_{\vk} 
\equiv 
\sqrt{\epsilon_{\vk \sigma}^2 
+ [\Delta_{\vq} \gamma_{\alpha}(\vk)]^2}$, 
$E_{\vk \sigma} \equiv E_{\vk} + \sigma h$, 
and $h \equiv \mu_0 |{\vector H}|$~\cite{Shi94,Shi97a,Shi97b,Shi00b}. 
By an approximation of the standard weak coupling theory, 
we obtain 
\def\eqgapeqweakcoupling{(2)}
$$
     \begin{array}{rcl}
     \lefteqn{ \dps{ \log \frac{T_{\rm c}^{(0)}}{T} }
     = \dps{ 
       \int_0^{\infty} \hsp{-1} \d y \, 
       \int_{-\pi}^{\pi} \frac{\d \theta}{2 \pi} 
         \frac{\rho_{\alpha}(0,\theta)}{N_{\alpha}(0)} 
     }} \\[12pt]
     && \dps{ 
         \times 
         \sinh^2 \frac{\beta \zeta}{2}
         \frac{\tanh y }
              {y \, ( \cosh^2 y + \sinh^2 \frac{\beta \zeta}{2} )} , 
     }
     \end{array}
     \eqno\eqgapeqweakcoupling
     $$
where 
\def\eqNalpharhoalpha{(3)}
$$
     \begin{array}{rcl} 
     \zeta 
       & \equiv & \dps{ 
         h \, ( \frac{{\vector v}_{\rm F} \cdot \vq}{2 h} - 1 ) , 
     }\\[8pt] 
     \rho_{\alpha}(\epsilon,\theta) 
       & = & 
         \rho(\epsilon,\theta) \, 
         {\Bigl [} 
         [\gamma_{\alpha}(\vk)]^2 
         {\Bigr ]}_{|\vk| = \kF(\theta)} , 
     \\[8pt] 
     N_{\alpha}(\epsilon) 
       & = & \dps{ 
         \int_{-\pi}^{\pi} \frac{\d \theta}{2\pi} 
           \rho_{\alpha}(\epsilon,\theta) , 
     }
     \end{array}
     \eqno\eqNalpharhoalpha
     $$
and ${\vector v}_{\rm F}$ is the Fermi velocity. 
Here, $\rho(\epsilon,\theta)$ is the angle dependent density of states 
which satisfies 
\def\eqrhoalpha{(4)}
$$
     \frac{1}{N} \sum_{\vk} F(\epsilon_{\vk}) 
     = \int \d \epsilon \, 
       \int_{-\pi}^{\pi} \frac{\d \theta}{2 \pi} 
         \rho(\epsilon,\theta) F(\epsilon ) 
     \eqno\eqrhoalpha
     $$
for any function $F(\epsilon_{\vk})$, 
and $\kF(\theta)$ is the magnitude of the Fermi momentum in the 
direction of $\theta$, which is the angle measured from the $k_x$-axis. 
By solving equation eq.~{\eqgapeqweakcoupling} with $\vq$ fixed, 
we obtain a transition temperature $T(h,\vq)$ 
or a critical field $h(T,\vq)$ for the value of $\vq$. 
The final result of the transition tempreature is given by 
\def\eqTc{(5)}
$$
     \Tc(h) = \max_{\vq} (T(h,\vq)) 
     \eqno\eqTc
     $$

As we mentioned above, we use the square lattice tight-binding model 
\def\eqepsilon{(6)}
$$
     \epsilon_{\vk} = - 2 t \, (\cos k_x + \cos k_y) - \mu 
     \eqno\eqepsilon
     $$
to express various anisotropic Fermi-surfaces. 
The cylindrically symmetric Fermi-surface is obtained 
in the limit of $\mu \rightarrow - 4t$, 
whereas a square Fermi-surface is obtained 
in the limit of $\mu \rightarrow 0$. 
Figure 1 shows the shape of the Fermi-surfaces of 
$\mu= - 0.5t$, $-1.8t$ and $-3t$. 
In this model, $\zeta$ is written as 
$
     \zeta = h \, ( \qbar \, x(\theta, \varphi_{\vq}) - 1) 
     $
with $\qbar = t q /h$ and 
$
     x(\theta, \varphi_{\vq}) 
     =   
       [   \cos \varphi_{\vq} \sin (k_x) 
         + \sin \varphi_{\vq} \sin (k_y) ]_{|\vk| = \kF(\theta)} , 
     $
where $\varphi_{\vq}$ is the angle between $\vq$ and $k_x$-axis. 
We examine $d_{x^2-y^2}$-wave pairing 
\def\eqgammad{(7)}
$$
     \gamma_{d_{x^2-y^2}}(\vk) 
     = [\cos k_x - \cos k_y]_{|\vk|=\kF(\theta)} , 
     \eqno\eqgammad
     $$
which is more favorable than $d_{xy}$-wave pairing, 
because of 
the peak in the angle dependent density of state $\rho(\epsilon,\theta)$ 
near the saddle points of the electron dispersion relation.

\vspace{\baselineskip}

\begin{figure}[htb]
\begin{center}
\leavevmode \epsfxsize=6cm  
\epsfbox{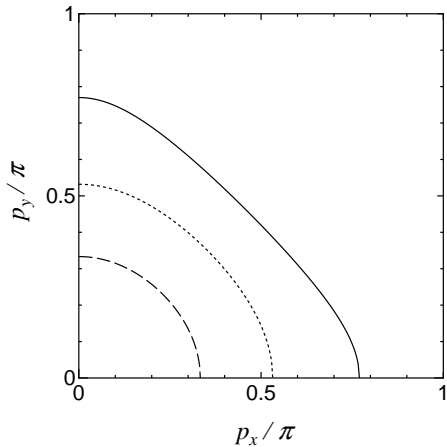}
\end{center}
\caption{
The solid, dotted and dashed lines show the Fermi-surfaces at 
$\mu = -0.5t$, $-1.8t$ and $3t$, respectively. 
The lattice constant is taken as unity. 
} 
\label{fig:fs}
\end{figure}

\begin{figure}[htb]
\begin{center}
\leavevmode \epsfxsize=6cm  
\epsfbox{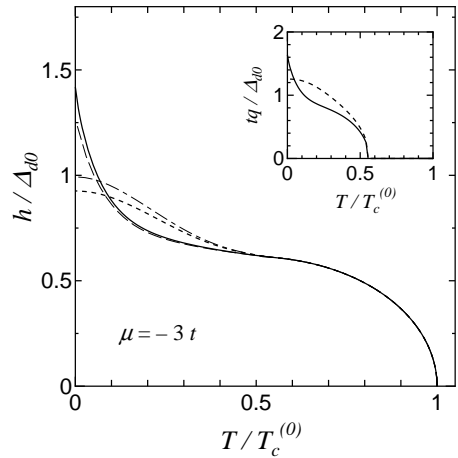}
\end{center}
\caption{
The solid and dotted lines show the critical fields 
for $\varphi_{\vq} = 0$ and $\pi/4$, respectively, 
when $\mu = -3t$, i.e., $n \approx 0.17$. 
In the inset, the solid and dotted lines show the temperature 
dependences of $q= |\vq|$ along the transition lines, 
for $\varphi_{\vq} = 0$ and $\pi/4$, respectively. 
The dashed and dot-dashed lines show the results of 
cylindrical symmetric Fermi-surface, (i.e., $\mu \rightarrow -4t$), 
for $\varphi_{\vq} = 0$ and $\pi/4$, respectively. 
}
\label{fig:mu3}
\end{figure}

Figure~\ref{fig:mu3} shows the critical field for 
$\mu = -3 t$, i.e., $n \approx 0.17$. 
In the vertical axis, $h = \mu_0 |\vector H|$ 
is scaled by $\Delta_{d0} \equiv 2 \omega_{\rm D} \exp(-1/VN_d(0))$, 
which is rather different from the zero field BCS gap for anisotropic 
pairing~\cite{Shi99}. 
As shown in Fig.~\ref{fig:fs}, the Fermi-surface has a nearly 
cylindrical shape at this value of $\mu$. 
The behavior of the critical field is similar to 
that for the cylindrically symmetric Fermi-surface, 
which corresponds to $\mu \rightarrow -4t$. 
The optimum direction of $\vq$ of the FFLO state changes 
at a temperature $T \approx 0.1 T_{\rm c}^{(0)}$, 
where the kink appears. 
Weak anisotropy does not change this behavior. 
It is confirmed by numerical calculations 
that the other directions $\vq$ 
than $\varphi_{\vq} = 0$ and $\pi/4$ 
give lower critical fields than the highest one.

\vspace{\baselineskip}

\begin{figure}[htb]
\begin{center}
\leavevmode \epsfxsize=6cm  
\epsfbox{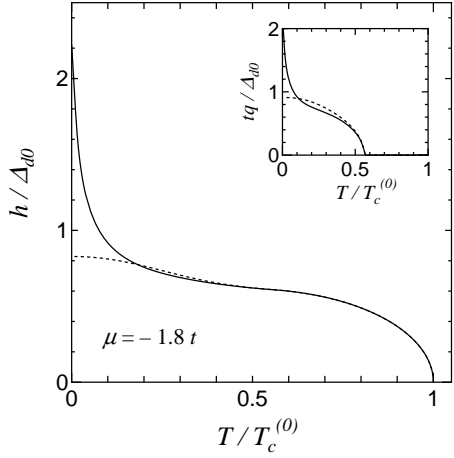}
\end{center}
\caption{
The solid and dotted lines show the critical fields 
for $\varphi_{\vq} = 0$ and $\pi/4$, respectively, 
when $\mu = -1.8t$, i.e., $n \approx 0.41$. 
In the inset, the solid and dotted lines show the temperature dependences 
of $q= |\vq|$ along the transition line 
for $\varphi_{\vq} = 0$ and $\pi/4$, respectively. 
} 
\label{fig:mu18}
\end{figure}

\begin{figure}[htb]
\begin{center}
\leavevmode \epsfxsize=6cm  
\epsfbox{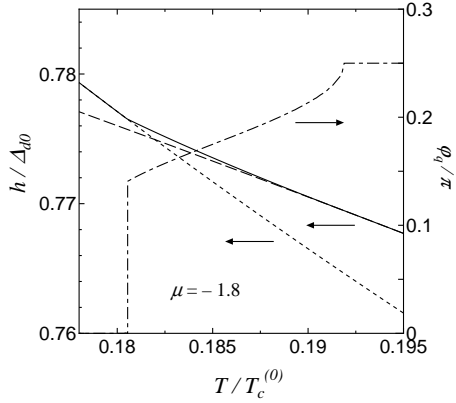}
\end{center}
\caption{
Behaviors of the critical fields and $q = |\vq|$ near the kink, 
where the lines of $\varphi_{\vq} = 0$ and $\pi/4$ cross, 
when $\mu = - 1.8t$. 
The dotted and dashed lines show the critical fields of 
$\varphi_{\vq} = 0$ and $\pi/4$, respectively. 
The solid line shows the final result of the critical field obtained 
by optimizing both the direction and the magnitude of $\vq$. 
The dot-dashed line shows the optimum $\varphi_{\vq}$. 
} 
\label{fig:mu18kakudai}
\end{figure}

Figure~\ref{fig:mu18} shows the results for $\mu = - 1.8t$, 
i.e., $n \approx 0.41$. 
As shown in our previous paper~\cite{Shi99}, the critical field 
increases markedly near $\mu = -2t$ due to an effect of the structure of 
the Fermi-surface in the present model. 
Therefore, for $\mu = -1.8t$, the critical field is very large at $T = 0$. 
It is found that the enhancement occurs especially at low temperatures, 
where $\varphi_{\vq} = 0$ is the optimum direction. 
At rather higher temperatures, $\varphi_{\vq} = \pi/4$ is the optimum 
direction, which gives critical fields slightly higher than that of 
$\varphi_{\vq} = 0$. 
As shown in Fig.~\ref{fig:mu18kakudai}, 
other directions are optimum near the kink, where the curves of 
$\varphi_{\vq} = 0$ and $\pi/4$ cross, 
although the differences in the magnitudes of the critical fields 
are very small there. 
Such behavior does not occur for $\mu = -3t$ and $\mu \rightarrow -4t$.

\begin{figure}[htb]
\begin{center}
\leavevmode \epsfxsize=6cm  
\epsfbox{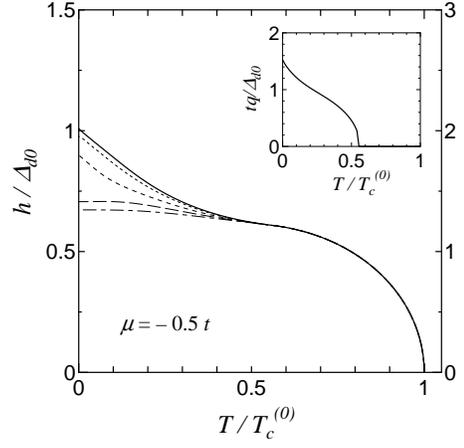}
\end{center}
\caption{
The solid, dotted, short dashed, dashed and dot-dashed lines 
show the critical fields 
for $\varphi_{\vq} = 0$, $\pi/20$, $\pi/10$, $3\pi/20$, and $\pi/4$, 
respectively, when $\mu = -0.5t$, i.e., $n \approx 0.77$. 
The inset shows the temperature dependence of $q= |\vq|$ 
along the transition line. 
} 
\label{fig:mu05}
\end{figure}

Figure~\ref{fig:mu05} shows the critical field for $\mu = - 0.5t$, 
i.e., $n \approx 0.77$, 
where the Fermi-surface has a nearly square shape. 
It is found that the optimum $\vq$ is oriented to the direction of 
the crystal axis ($\varphi_{\vq} = 0$) 
in a whole temperature region, 
and thus the kink does not occur.

\vspace{\baselineskip}

In conclusion, we have studied the upper critical field of the 
superconductor on the square lattice in parallel magnetic fields. 
In particular, we have examined anisotropy effects of Fermi-surface 
on the temperature dependence of the upper critical field. 
By changing the hole concentration, we have controlled the shape of 
the Fermi-surface. 
The results are summarized as follows: 
(1) The kink of the critical field as a function of the temperature 
disappears due to an anisotropy of the Fermi-surface of some extent 
(Fig.~\ref{fig:mu05}); 
(2) When the anisotropy is weak, the kink remains (Fig.~\ref{fig:mu3}); 
(3) Depending on the shape of the Fermi-surface and the temperature, 
the optimum $\vq$ is oriented to other directions than 
the symmetric directions 
$\varphi_{\vq} = 0$ and $\pi/4$ (Fig.~\ref{fig:mu18kakudai}); 
(4) The critical field increases rapidly 
especially at low temperatures (Fig.~\ref{fig:mu18}) 
near $\mu \approx - 2t$, 
where the nesting is most effective~\cite{Shi99}. 
From the result (1), 
the absence of the kink in the experimental data which exhibits 
the upturn at low temperatures does not exclude the possibility of 
the FFLO state of anisotropic singlet pairing.

When the orbital pair-breaking effect remains to some extent, 
for example, for large inter-layer coupling, 
the present result must be taken in another way. 
In this case, the direction of $\vq$ coincides with the direction of 
the vortex lines for the coexistence of the vortex state and 
the FFLO state. 
Hence, $\varphi_{\vq}$ must be regarded as the direction of 
the applied magnetic field in Figures~\ref{fig:mu3}, \ref{fig:mu18} 
and \ref{fig:mu05}. 
The temperature dependence of the upper critical field changes 
qualitatively as shown in those figures, 
by change of the direction of the applied field 
between $(1,0,0)$ and $(1,1,0)$ directions, 
when the vortex state coexists.



\end{document}